\input{aipcheck}

\documentclass[
    ,final            
  ]
  {aipproc}

\layoutstyle{6x9}

\begin{document}

\title{BL Lac Contribution to the Extragalactic Gamma-Ray Background}

\author{Tanja M. Kneiske}{
  address={Universitaet Wuerzburg, Am Hubland, 97074 Wuerzburg, Germany}
}

\author{Karl Mannheim}{
  address={Universitaet Wuerzburg, Am Hubland, 97074 Wuerzburg, Germany}
}

\begin{abstract}
Very high energy gamma-rays ($E_\gamma >$ 20 GeV) from blazars
traversing cosmological distances through the metagalactic radiation
field can convert to electron-positron pairs in photon-photon collisions.
The converted gamma rays initiate electromagnetic cascades
driven by inverse-Compton scattering off the microwave background photons.
The cascades shift the injected gamma ray spectrum to MeV-GeV energies. Randomly
oriented magnetic fields rapidly isotropize the secondary electron-positron beams
resulting from the beamed blazar gamma ray emission, leading to faint
gamma-ray halos.  Using a model
for the time-dependent metagalactic radiation  field consistent with all
currently available far-infrared-to-optical data, we compute (i) the
expected gamma-ray attenuation in blazar spectra, and (ii) the
cascade contribution from faint, unresolved blazar to the extragalactic
gamma-ray  background as measured by EGRET, assuming a generic emitted
spectrum extending to an energy of 10 TeV.  The latter cascade
contribution to the EGRET background is fed by the assumed >20 GeV
emission from the hitherto undiscovered sources, and we estimate their
dN-dz distribution taking into account that the nearby (z<0.2) fraction of
these sources must be consistent with the known (low) numbers of sources above
300 GeV.

\end{abstract}

\maketitle


\section{Introduction}
An isotropic, diffuse
background radiation presumably due to faint, unresolved extragalactic sources
has been observed in nearly all energy bands.
The confirmation of an extragalactic gamma-ray background by EGRET (Energetic
Gamma-Ray Experiment Telescope) on board the Compton Gamma Ray Observatory
has extended the spectrum up to
an energy of $\sim$50~GeV.
A first analysis of the data resulted in
a total flux of $(1.45\pm 0.05) \cdot 10^{-5}$ photons cm$^{-2}$ s$^{-1}$ 
sr$^{-1}$ above 100~MeV and a spectrum which could be fitted by
a power law with an spectra index of $-2.1\pm 0.03$ \cite{lit:sreekumar}. 
These values are strongly dependent on the
foreground emission model which is subtracted from the observed
intensity to obtain the extragalactic residual \cite{lit:hunter}.
Since using the old foreground model, a residual GeV halo remained after subtraction (in
addition to the isotropic extragalactic background), the foreground model had to be
improved.  This lead to a new analysis of the EGRET data, and a new result for the
extragalactic background spectrum, now showing a dip a GeV energies
and an overall weaker intensity of
$(1.14\pm 0.12) \cdot 10^{-5}$ photons cm$^{-2}$ s$^{-1}$ sr$^{-1}$ 
\cite{lit:strong}.
This new result can help us to understand the origin of the extragalactic
background radiation.

Since EGRET detected a large number of extragalactic
gamma-ray sources belonging to the blazar class of AGN, a reasonable assumption is that the gamma background
is produced by unresolved AGN.
Using a gamma-ray luminosity function from EGRET data
\cite{lit:chiang}Chiang \& Mukherjee (1998) came to the result that only
25\% to 50\% of the gamma background could be explained by blazars.
\cite{lit_stecker96} were able to explain 100\% of the background, but were facing a problem
with the deficit of observed faint, nearby blazars.

The new idea which will be presented in this paper is to extend the
existing models by assuming a population of BL Lacs with a spectral
energy distribution such that their flux at EGRET energies is too low
to be generally detected, while their very high energy gamma ray flux is strong.
Since most of these sources are at redshifts high enough for pair attenuation to
take place, a significant part of their VHE emission is reprocessed by cascades
contributing to the diffuse background, but not to the single source counts.

During the paper we use a Hubble constant of $H_0=71$km s$^{-1}$ Mpc$^{-1}$ and
a flat universe with the cosmological parameters $\Omega$=0.3 and $\Omega_\Lambda$=0.7.

\begin{figure}
  \includegraphics[height=.3\textheight]{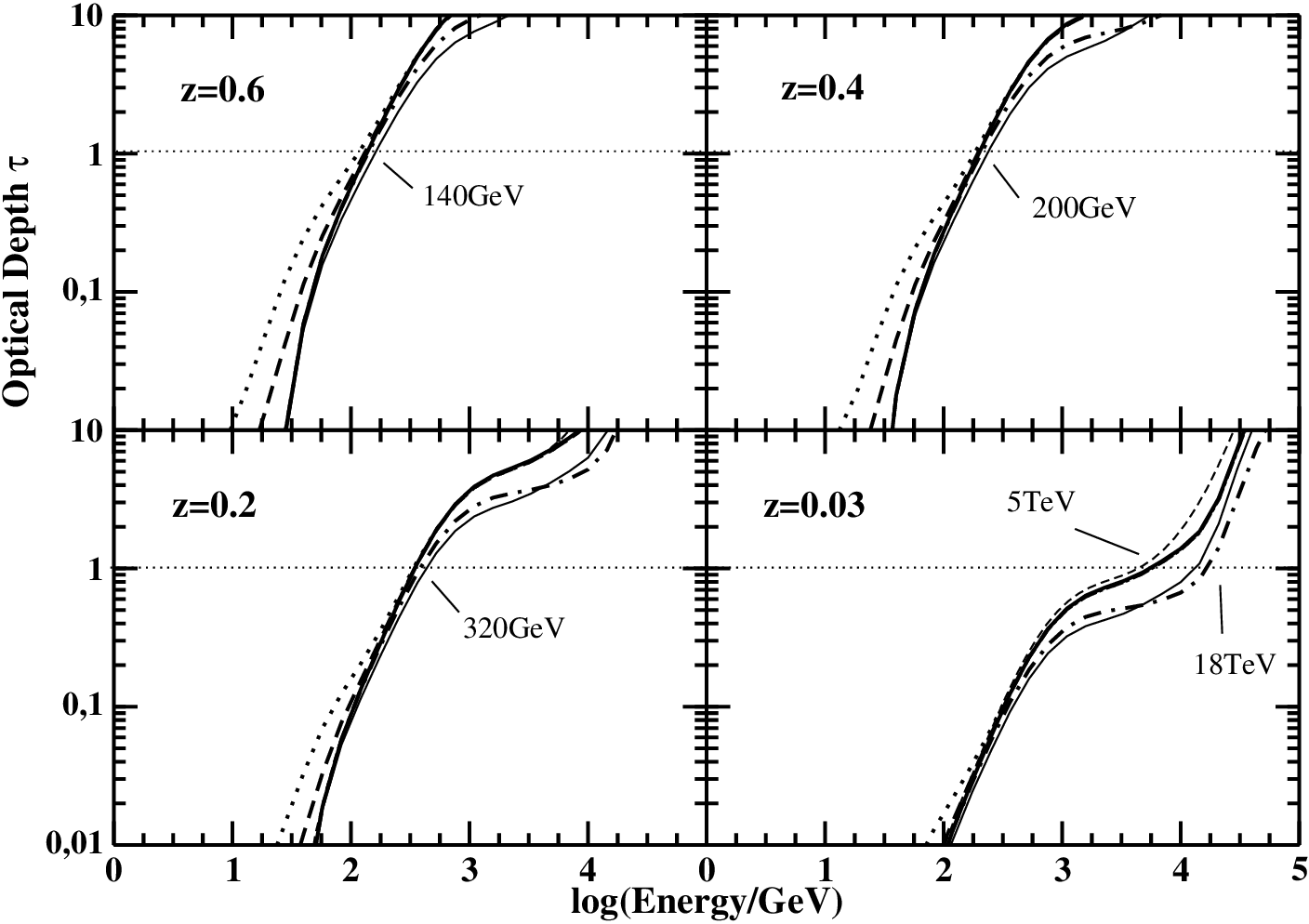}
  \caption{Pair attenuation optical depth for various redshifts and MRF models.
  The labeling of the line styles are explained in 
  \cite{lit:kneiske2}.
  The crossing point with
  the line $\tau=1$ defines the exponential cutoff energy.}
  \label{fig:Tau}
\end{figure}

\section{Gamma-Ray Background}

If the gamma-ray background is produced by unresolved sources, 
it can be described by

\begin{equation}
F_{E_\gamma} = \frac{1}{\Omega} \int^{z_m}_0 dz \frac{dV}{dz}  
\int^{\infty}_{L_{\rm min}} \frac{dN}{dV dL}
 F_{E_\gamma}(z, L) dL, \ \ \ \
 \label{eq:gammaback}
\end{equation}
with $\Omega$ the solid angle coverage of the survey 
($\Omega_{\rm EGRET}=10.4$), $\frac{dV}{dz}$ the volume element, 
$L_{min}$ is the luminosity of the weakest source, $\frac{dN}{dV dL}$
the luminosity function and $F_{E_\gamma}(z_q, L)$ the flux of the 
gamma sources, depending on their luminosity and redshift.
The luminosity function of resolved EGRET sources, extended to the faint end
has been computed by \cite{lit:chiang}.
We used their model changing only the spectral index from $\alpha=2.1$ to
$\alpha=2.3$. The new spectral index was determined by fitting
the reanalyzed EGRET data at $ <2.0$~GeV.

The remaining excess of the measured gamma-ray background we ascribe to high-energy
peaked blazars belonging to the HBL and ExBL classes (defined by \cite{lit:Ghisellini}.
We calculate the flux from this sources using equation~\ref{eq:gammaback}.
The spectral energy distribution of these sources between 100 MeV and 10 TeV and
their luminosity function (LF) is
poorly known, and we have
to make some theoretical assumptions for them which are described
the next sections.

\begin{figure}
  \includegraphics[height=.3\textheight]{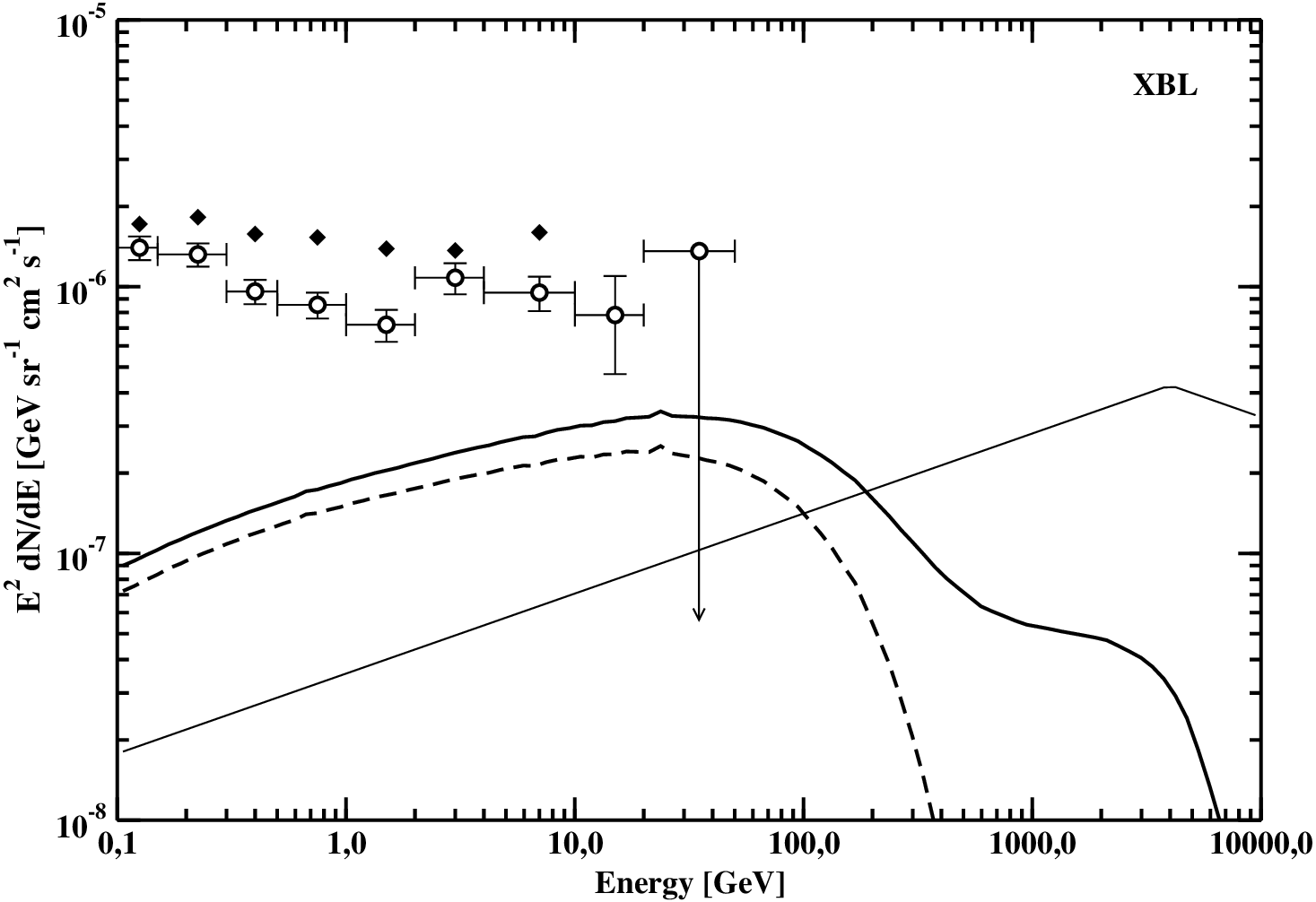}
  \caption{Spectrum of the extragalactic gamma-ray background (open circles: Sreekumar (1998),
  filled diamonds: Strong et al. (2004)).
  The contribution of the HBL component (thick solid line) is compared
  with the spectrum without any absorption and reemission (thin solid line)
  and the contribution of the secondary photons only (dashed line)}
  \label{fig:gammabackHBL}
\end{figure}

\section{Template Spectra}

A number of extragalactic gamma-ray sources have been 
detected with imaging air-Cherenkov telescopes (Table 6, \cite{lit:horan}).
Four of them (with redshifts $z=0.03, 0.03, 0.129, 0.048$)
were bright enough to resolve their spectra in the TeV energy band.
The observed spectra are presumably modified by gamma ray attenuation, i.e.
\begin{equation}
F_{\rm obs}(E)=F_{\rm int}(E)\exp[-\tau_{\gamma\gamma}(E,z)]
\end{equation}
where $\tau_{\gamma\gamma}(E,z)$ is the optical depth for gamma-rays
(Fig.~\ref{fig:Tau}).
We used various model parameters for the metagalactic radiation field (MRF) to bracket the
range of the un-absorbed (intrinsic) spectra.

Depending on the model of the MRF the intrinsic spectra show turnovers
or broad maxima around a few TeV. The intrinsic spectrum of H1426+428
which has a larger redshift (z=0.129) could have a maximum at 10~TeV or
higher. We use a mean of the spectra of Mkn501, Mkn421 and ES1959+650
as a template for the HBL-type sources and
a spectrum like H1426+428 as the template spectrum for the ExBL-types.
Each template spectrum is modeled using
two power laws. The parameters are two spectral
indices and the location of the maximum. 
For HBL the spectral index at low energies is $\alpha=1.7$ and
at high energies $\alpha=2.3$ with a maximum at 4~TeV.
The spectral index of the ExBLs is $\alpha=1.2$ with a maximum at 10~TeV.

\begin{figure}
  \includegraphics[height=.3\textheight]{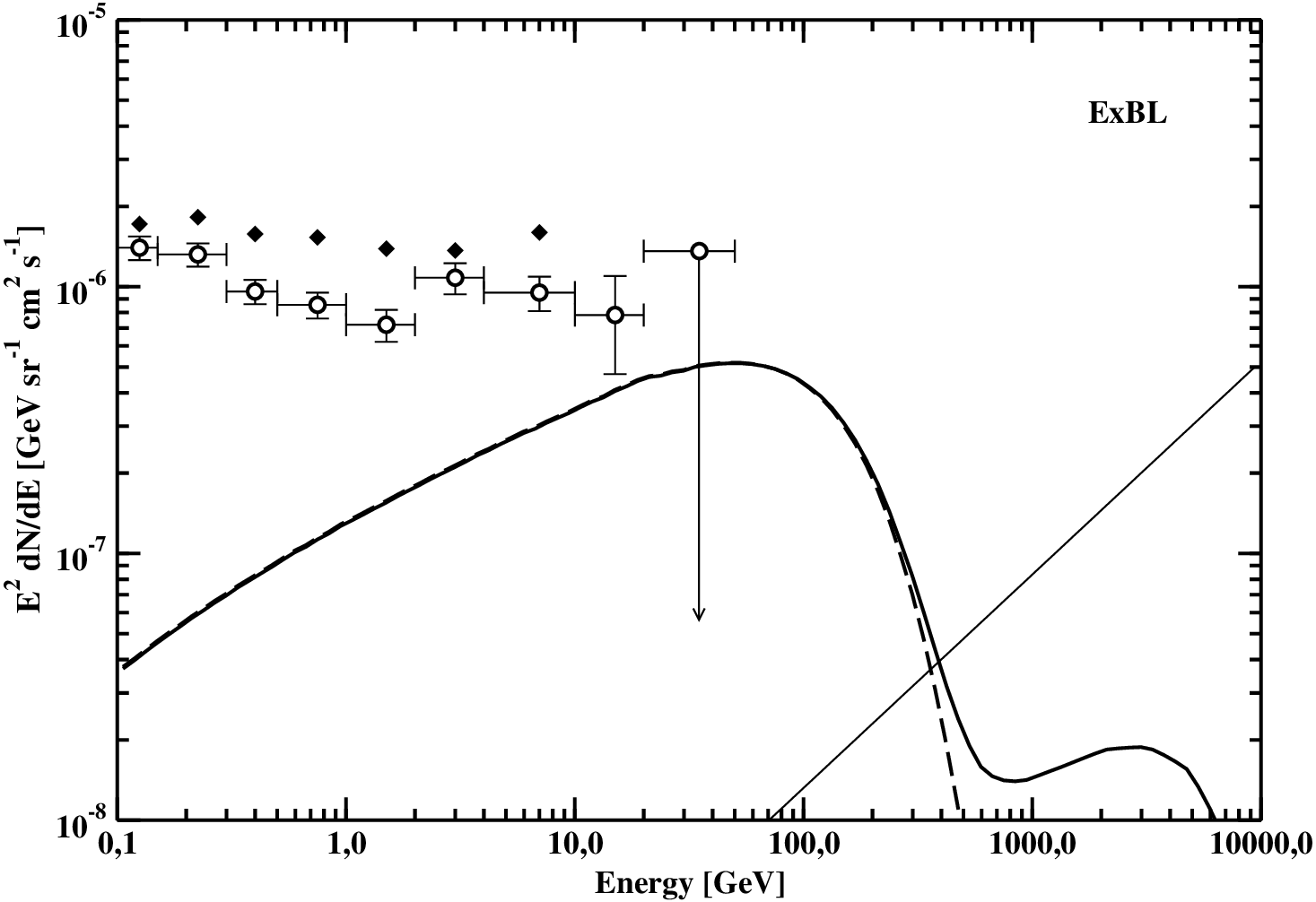}
  \caption{Spectrum of the extragalactic gamma-ray background.
  The contribution of the ExBL component (thick solid line) is compared
  with the spectrum without any absorption and reemission (thin solid line)
  and the contribution of the secondary photons only (dashed line)}
  \label{fig:gammabackExBL}
\end{figure}

The absorption of the primary photons is calculated using the MRF model 
presented in \cite{lit:kneiske1}, and the reemission is calculated
using the radiative transfer equation employing an inverse-Compton emission term
due to scattering off the microwave background.
The flux from a population of
gamma-ray sources contributing
to the gamma-ray background is the sum of the primary and the secondary
flux
$F_{E_\gamma}(z_q, L)=F^{\rm p}_{E_\gamma}(z_q, L)+F^{\rm s}_{E_\gamma}(z_q, L)$.

\section{TeV-Luminosity Function}

Although the secondary cascade contribution is likely to be isotropic, and therefore much fainter
than the beamed primary gamma rays, we ignore this effect which is compensated by the correspondingly
larger number of hosts which contribute to the extragalactic background light.  By doing so, we
must use the LF of the parent population of the secondary emission, e.g. the LF of XBLs and ExBLs.
HBL and ExBL are x-ray selected BL Lacs showing indications of correlated x-ray/gamma ray emission.
We will use
x-ray observations to develop a TeV-luminosity function.
Using the ROSAT-All-Sky-Survey, \cite{lit:Bade}
and \cite{lit:laurent} could
develop a luminosity function with a  maximum in the number of sources at z=0.2.
\cite{lit:rector} and \cite{lit:caccianiga} found a maximum of
sources around z=0.3 using a sample of the Einstein-Medium-Sensitivity Survey
and the Radio-Emitting-X-Ray-Sources catalog, respectively.
We will instead use the results of
\cite{lit:Beckmann} who combined all the available data.
To obtain a relation between the gamma-ray flux and x-ray flux we used
the calculations from \cite{lit:costaghiss}. They presented
for 33 BL Lacs multi-wavelength spectra using a theoretical SSC-model
and various observations. From their results we fitted a relation between
the luminosity at 1~keV $L_{(1~\rm keV)}$ and the gamma-ray luminosity above 0.3~TeV $L_{(\rm TeV)}$

\begin{equation}
\label{eq:LTeVLx}
L_{\rm TeV}=2.6535 \cdot 10^{-4} L_{(1~\rm keV)}^{0.15781} \ \ \
 [10^{48}\mathrm{erg \ s}^{-1} ].
\end{equation}

The TeV-luminosity function can be written as broken power law
$\frac{dN}{dV}(dL_{0, \rm TeV}) \propto \left(L_{0, \rm TeV}\right)^{\alpha_{\rm LF}}$
with $\alpha_{\rm LF}=-0.9$ for $L_{0, \rm TeV} \le L_{\rm B}$ and $\alpha_{\rm LF}=-1.4$ 
for $L_{0, \rm TeV} > L_B$ with $L_B$ break luminosity.
The evolution can be described by
\begin{equation} 
\rho(z) \propto (1+z)^{\alpha_\rho}    .
\label{eq:dNdz_bllac}
\end{equation}
with a steep rise of $\alpha_{\rho}=10$ for $z \le 0.15$ and $\alpha_{\rho}=-3$ for
$z > 0.15$ (values fitted from the distribution shown in Beckmann et al. 2003).

The maximum and the minimum of the luminosity function has been calculated
from the maximum and minimum of the x-ray LF. The absolute value of the
luminosity function has been chosen to match the EGRET gamma-ray background
data. The bright ($L_{(\rm TeV)}>2.7\cdot 10^{-4}$ $10^{-48}$erg s$^{-1}$) sources are defined as ExBLs while the faint end
of the LF is assumed to represent the HBLs. The number density is comparable
with the observed EGRET blazar number density.

\section{Discussion}

\begin{figure}
  \includegraphics[height=.3\textheight]{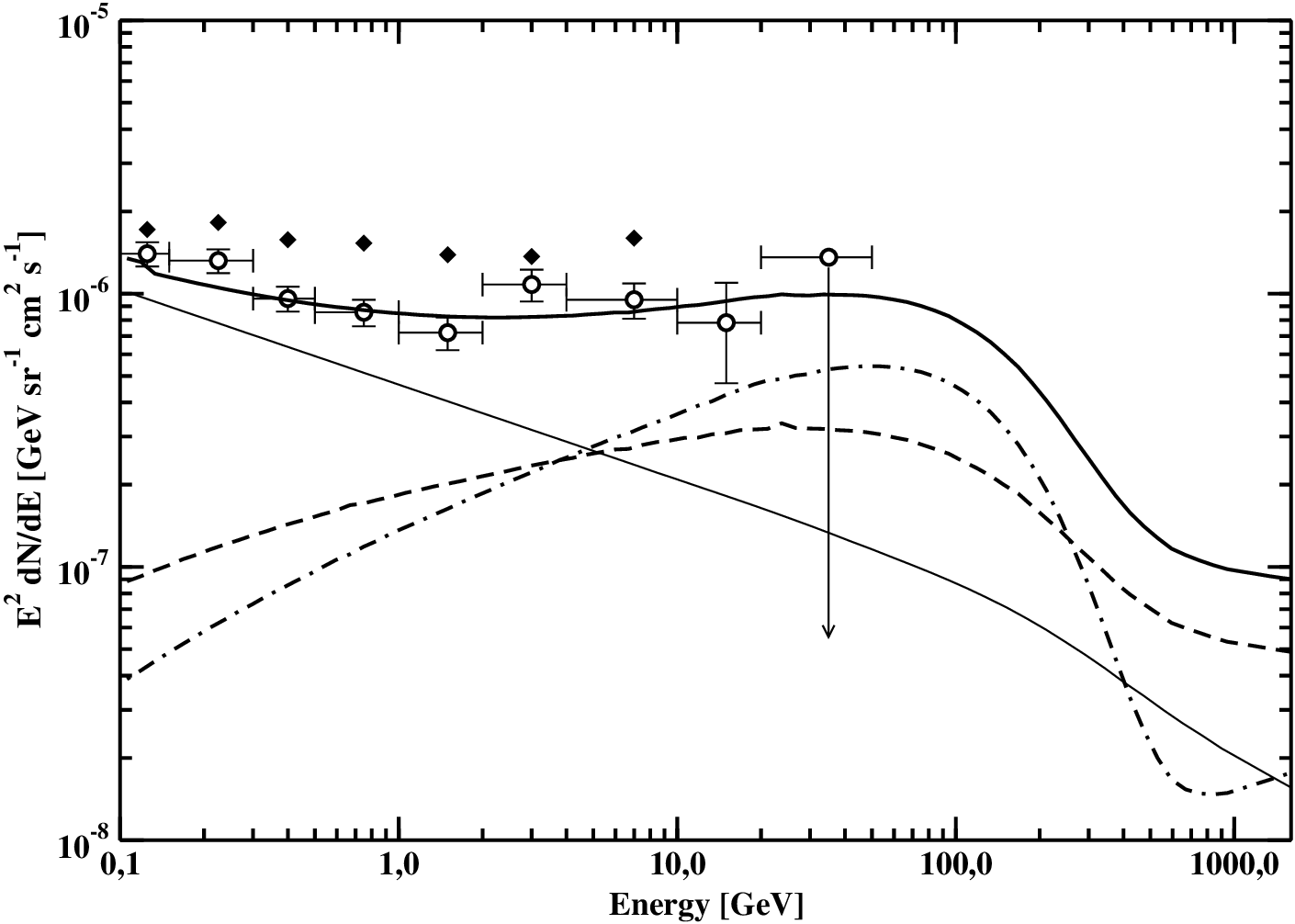}
  \caption{Spectrum of the extragalactic gamma-ray background.
  The total spectrum is produced by three components:
  EGRET sources with a spectral index of 1.3 (thin solid line), 
  HBL (dotted-dashed line) and ExBL (dashed line). For all three
  component the effect of extragalactic absorption and reemission
  via inverse Compton scattering is taken into account.}
  \label{fig:gammaback}
\end{figure}

The results for the various contributions to the extragalactic gamma-ray
background are shown in Figure~\ref{fig:gammaback}.
The thin solid line denotes the EGRET blazar contribution.
Due to the new spectral index and the reanalyzed EGRET data the unresolved
EGRET blazars now produce about 75-80\% of the background flux.
The dashed line represents the background flux made due to the HBL population, while
the dot-dashed line describes the flux corresponding to the ExBL contribution.
Comparing the total flux as a sum of the three contributions and
the EGRET data, the agreement is acceptable.
As can be seen in Fig.~\ref{fig:gammabackHBL} and
Fig.~\ref{fig:gammabackExBL} the primary flux of the BL Lac population produces only a small contribution
in the EGRET energy range. The secondary photons can contribute about
20\% to the gamma-ray background.
Although the number of ExBL is less then 10~\% of the total number
of BL Lac objects the flux is of the same order of magnitude as the
HBL flux. The adopted values of the spectral index and
the maximum at 10 TeV have a large influence on the secondary photon contribution.

The number density of BL Lacs in our model is comparable with that of the EGRET blazars.
Nevertheless,  only 7 HBLs have been observed with Imaging Air Cherenkov Telescopes above 300 GeV
(in the published records).  The reason for this comparatively small number is the effect of
pair attenuation and the limited sensitivity of the Cherenkov telescopes.
E.g., for AGN at a redshift of 0.2 photons with energies $>300$~GeV are
undergoing the pair production process. With Cherenkov Telescopes of the
Whipple tpye ($E_{\rm thr}\approx 300$, $F_{\rm lim} \approx 10^{-11}$)
still most sources within this redshift range are too faint to be detected in a typical 10-50 h observation campaign.
The observational constraint of small zenith angles further reduces the number of observable
sources number by a factor of roughly $\sim 1/4$.
Assuming the number density to fit the gamma-ray background data
a telescope of the Whipple type would only be able to observe $\approx$ 18 BL Lacs from this population.

\begin{figure}
  \includegraphics[height=.3\textheight]{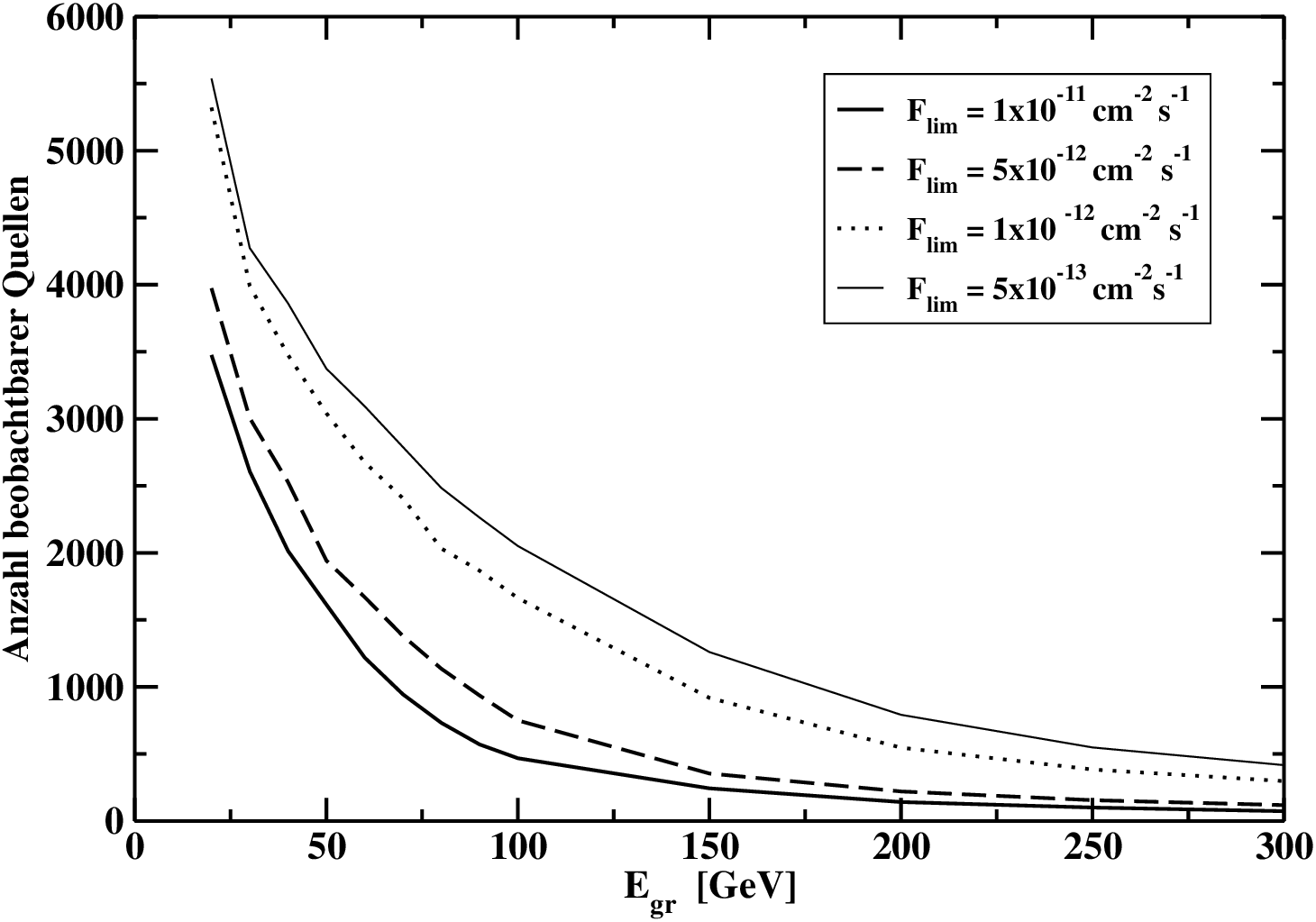}
  \caption{Number of observable BL Lacs as a function of energy threshold and
  flux limit of Cherenkov Telescopes.}
  \label{fig:dNdz}
\end{figure}

Our assumptions can be tested by
the next generation of Cherenkov Telescopes.
Fig.~\ref{fig:dNdz} shows the number of BL Lacs depending on the
flux limit and the threshold energy of a telescope.
A telescope with a threshold energy of 50~GeV and a flux limit of
$F_{\mathrm lim} \approx 10^{-11}$ should in principle be able to observe about 25\% of 1500
sources.

\begin{theacknowledgments}
This research was gratefully supported by the BMB+f under grant 05AM9MGA.
\end{theacknowledgments}

\end{document}